\documentstyle[12pt]{article}
\begin{document}

\begin{center}

\vspace*{0.5in}

\baselineskip=26pt
{\LARGE
MY ENCOUNTERS --- AS A PHYSICIST --- WITH MATHEMATICS}

\vskip1in

\baselineskip=20pt
{\Large
R. Jackiw \\[1ex]
M.I.T., Cambridge~ MA}

\bigskip

{\large
MIT-CTP \#2371 ~~ --- ~~ HEP-TH \#9410151}

\vfill

\baselineskip=17pt
{\large
25th anniversary and new building dedication \\
Centre de Recherches Math\'{e}matiques \\
Montr\'{e}al, Canada, October 1994}

\vspace*{0.5in}

\end{center}

\newpage

{\large
\baselineskip=24pt
\parskip=1.5ex

It is a pleasure for me --- as a physicist --- to speak at the
twenty-fifth anniversary of your mathematical institute, when also a
new building is being dedicated.  Such an occasion calls for accolades
of appreciation, both to the profession and to its practitioners, and
over the years many bouquets have been offered to mathematics.

For example, Gauss was quite pleased with his line of work, stating that

\begin{quote}
``Mathematics is the queen of the sciences.''
\end{quote}
\vspace{-3ex}\rightline{(Gauss)}\vspace{2ex}\noindent
Even an outsider, the psychologist Havelock Ellis, evaluated
mathematicians as having

\begin{quote}
``\dots
reached the highest rung on the ladder of human thought''
\end{quote}
\vspace{-3ex}\rightline{(Ellis)}\vspace{2ex}\noindent
To be sure, there are also dissenters, for example Plato:

\begin{quote}
``I have hardly ever known a mathematician who was capable of reasoning.''
\end{quote}
\vspace{-3ex}\rightline{(Plato)}\vspace{2ex}\noindent

Physicists' opinion lies between these extremes.  Well known is
Wig\-ner's appreciation of the

%%% \samepage{
\begin{quote}
``unreasonable effectiveness of mathematics'' \nobreak
\end{quote}
\nobreak\nobreak
\vspace{-3ex}\rightline{(Wigner)}
\medbreak
\noindent
A forceful position in favor of mathematics in physics was stated by Dirac:
%%% }

\begin{quote}
``The most powerful method of advance [in physics]
\dots
is to employ all the resources of pure mathematics in attempts to
perfect and generalize the mathematical formalism that forms the
existing basis of theoretical physics, and
\dots
to try to interpret the new mathematical features in terms of physical
entities.''
\end{quote}
\vspace{-3ex}\rightline{(Dirac)}\vspace{2ex}\noindent

The list of apt quotations can be extended to great length, and I
shall not attempt adding my own words.  Let me merely repeat that
mathematics is indeed good for us physicists, but also we are good for
mathematics by providing new ideas for mathematical research and by
finding fresh applications of old ideas.

These days there is intense cross fertilization between mathematics
and physics, specifically between geometry and field theory.  The
contact, first established through Einstein's general relativity,
surged again about two decades ago.  Some of my own research took
place at this new beginning, so I thought I would present here a
reminiscence, thereby providing a case history of a
physics-mathematics encounter.

By the early 1970's, quantum field theory was very much in favor with
theoretical physicists, but the quantized equations resisted solution.
It then occurred to many people that it would be worthwhile to ignore
the quantal nature of the fields, and to solve the equations as if
they describe non-linear, classical dynamical systems.  Interesting,
localized and non-dissipative solutions were found very quickly.
These were the kinks in one dimension --- relevant to physics on a
line, vortices in two dimensions --- in planar physics, Skyrmions and
magnetic mono\-poles in the three dimensions of our physical world;
collectively such solutions were called ``solitons,'' the name being
taken from applied mathematics.  Another class of solutions comprised
the ``instantons'' in four-dimensional space-time.

With colleagues at M.I.T., I addressed the question of how to extract
from these {\it classical\/} results information on the {\it
quantum\/} theory --- {\it i.e.\/}~we wanted to determine the quantum
meaning of classical fields.  Progress was made on this problem, and
at a certain stage Claudio Rebbi and I decided that we needed to study
both the linear small fluctuations about the non-linear soliton and
instanton field profiles, and also the coupling of other linear
systems, like fermions, to solitons and instantons.  Thus we were led
to linear eigenvalue equations and we realized that the
``zero-modes,'' corresponding to vanishing eigenvalues, contain
especially important information about the quantum physics.  The
zero-modes in the small fluctuation equations measure allowed
deformations of the soliton or instanton, while the number of these
modes gives the dimension of the moduli space for the solution of the
non-linear equation.  In the fermionic Dirac equation, the eigenvalues
measure energy, and positive-energy modes describe quantum particles,
negative-energy modes correspond to anti-particles, while zero-modes
--- when they exist --- signal a degeneracy that gives rise to
unexpected quantum numbers, {\it e.g.\/}~fractional fermion number.

Rebbi and I were delighted to find precisely such zero-modes, and to
establish their physical consequences.  But we were surprised that the
existence of these special solutions did not depend on the details of
the localized profiles in the background solitons and instantons;
rather only their large-distance behavior mattered.  The long-range
features of course characterize the topological properties of solitons
and instantons, so we began to suspect that the occurrence of
zero-modes was not an accident of our analysis, but a consequence of
having non-trivial topological backgrounds.

We wanted to find out what mathematicians knew about this.  At
M.I.T. all buildings are connected and the math department is in the
same structure as my work space.  However, locked doors as well as the
chemistry department intervene, so communication is obstructed.
Nevertheless, we walked the corridors of the mathematics offices, but
could not immediately find anyone who wanted to spend time
understanding our questions, and answering them in a way
comprehensible to non-specialists, to us physicists.  Shortly later we
met Barry Simon, who did not have specific information on our problem,
but suggested that work of Atiyah and Singer might be relevant.

Singer had temporarily moved from M.I.T. to Berkeley, but as it
happened my colleague and collaborator Goldstone knew Atiyah from
student days in Cambridge, England, and had information that he was
coming to visit his mathematics friends in Cambridge, Massachusetts.

So we arranged a meeting in my office.  We invited physicists who were
working on soliton-instanton questions, and we listened to Atiyah
explain how his index theorem with Singer counts instanton zero-modes,
and how their spectral flow theorem with Patodi is relevant to
fractional charge.  Learning that the four-dimensional index is given
by an integral over the curvature-form $F\/$, specifically by
$\displaystyle\int F \wedge F\/$, was especially thrilling to us since
the integrand, $F \wedge F\/$, had also arisen in the physics
literature as the anomalous divergence of the chiral fermion current,
thereby controlling neutral pion decay.  Evidently the chiral anomaly
and the index theorem are related; they had been elaborated in the
late 1960's at different ends of the same M.I.T. corridor, by people
working in ignorance of each other!

We appreciated very much Atiyah's efforts to make his presentation
understandable to us; still exchanging information was not easy.  One
young member of the audience impressed Atiyah, who encouraged the
fellow to speak because he seemed to understand, better than anyone
else, what Atiyah was saying.  That person was Witten; as all of you
know, he has continued to impress Atiyah and other mathematicians.

Soon thereafter, I was asked to review these exciting new results
about quantum field theory at a meeting of the American Physical
Society.  Since Singer was present, I yielded some of my time to him,
with the suggestion that he describe the mathematical connection.  But
a detailed presentation could not be fit in, so he merely eulogized
collaboration between mathematics and physics with the following ode.

\goodbreak\bigbreak
\begin{quote}
\hspace*{-1ex}
%%% \samepage{
\obeylines
``In this day and age
\nobreak
The physicist sage
%%% }
\nobreak
Writes page after page
On the current rage
The gauge

\vspace{3ex}

Mathematicians so blind
Follow slowly behind
With their clever minds
A theorem they'll find
Duly written and signed

\vspace{3ex}

But gauges have flaws
\nobreak
God hems and haws
\nobreak
As the curtain He draws
\nobreak
O'er His physical laws
It may be a lost cause''
\end{quote}
\vspace{-3ex}\rightline{(Singer)}\vspace{2ex}\noindent

Index theory also received a contribution from physics.  The
Atiyah-Singer theorem applies to even-dimensional spaces on which a
connection is defined.  However, physicists are also interested in
odd-dimensional spaces --- where one-dimensional kinks or
three-dimensional Skyrmions and mo\-no\-poles reside.  These
configurations can lead to zero modes, even in the absence of a gauge
connection.  So we asked the mathematicians about odd dimensions;
apparently nothing was known.  At that time I had a
mathematically-minded student, Costas Callias, and I asked him to
prove an odd-dimensional index theorem.  He succeeded and this further
prompted Bott and Seeley to publish a mathematical exegesis of the
result, immediately following Callias' paper in {\it Communications in
Mathematical Physics\/}.  Since then I have been happy to see the
``Callias index theorem'' used and cited.

%%%          put figure here
\begin{figure}[h]
\vspace{5.5in}
\caption{Front pieces of papers by Callias, Bott and Seeley.}
\end{figure}

\newpage

The two approaches to solving problems --- the explicit, goal-oriented
methods of the physicists and the general theorems of the
mathematicians --- are well illustrated by the determination of the
dimensionality for instanton moduli space: the n--instanton SU(2)
solution depends on 8n--3 parameters. This result appears in the same
issue of {\it Physics Letters\/}, once by Schwarz, who used the
Atiyah-Singer theorem, and a few pages later by Rebbi and me, who
solved differential equations to find explicitly 8n--3 zero-modes.

%%%          put figure here
\begin{figure}[h]
\vspace{5.5in}
\caption{Front pieces of papers by Schwarz, Jackiw and Rebbi.}
\end{figure}

\newpage

Gauge theories in general and instantons in particular continued to
interest mathematicians.  They established the topological properties
of the instanton moduli space and produced a construction --- but not
an explicit formula --- for the general solution.  The most general
{\it explicit\/} expression, which does not exhaust all the
parameters, was given by physicists.

Further developments on four-dimensional gauge fields led
mathematicians to Donaldson theory.  In three dimensions, the
Chern-Simons term --- another gauge structure first emphasized in the
physics literature --- has been related by mathematicians to knot
theory on manifolds with various topologies, while physicists applied
this term to experimental phenomena on the plane, like the Hall
effect.

These parallel investigations by physicists and mathematicians also
highlight our differences: physicists use mathematics as a language
for recording observations about physical systems, and this limits our
interest in the full range of mathematical possibility, which
fascinates mathematicians.  For example, the general instanton
solution does not appear to be physically relevant; only the original
one-instanton and the explicit, but limited, multi-instanton solutions
have illuminated physical theory.  Even the physicists' language need
not always be mathematical.  The fractional charge phenomenon, which
can be inferred from zero-modes or from spectral flow, was also
independently established by Su, Schrieffer and Heeger, who found a
physical realization in linear polyacetylene chains.  One of their
derivations uses the pictorial language of chemical bonds, where the
only mathematics consists of counting!

An analogy comes here to mind: the English language contains over
200,000 words and all of them interest the lexicographer; for
Shakespeare 20,000 words sufficed to express his ideas in plays and
sonnets; while Church\-ill used less than 2000 words in his
historically decisive speeches.  Physicists, like Churchill, achieve
their goals by effective use of a limited vocabulary.

The process of gaining knowledge goes through the same steps in
physics and mathematics: first there is the intuition/guess, then
follows the proposal/conjecture and finally comes the verification.
For the mathematician verification consists of constructing a proof,
establishing a theorem according to rules whose legitimacy evolves
slowly under the direction of the entire mathematics community.  But
the physicist verifies his ideas by finding a physical correlative:
neutral pion decay validates chiral anomalies, properties of solitons
in polyacetylene establish fractional charge.  The rules for giving a
proof are constantly and rapidly changing --- presuppositions can
become modified, experimental facts can evolve.

Because ``proof'' and ``theorem'' carry intellectual prestige and
pleasure, occasionally there are attempts to employ them in physics.
To my mind this is mostly futile and sterile.  For example, physicists
wanted very much to combine internal and space-time symmetries in a
non-trivial fashion and were not daunted by a proven ``impossibility
theorem.''  Rather the ``theorem'' was circumvented by the simple
device of replacing commutators with anti-commutators, by grading the
algebra, and supersymmetry was born, which now is also influencing
mathematics.  Similarly, when field-theory ``constructivists'' proved
the existence of quantum $\lambda \phi^4\/$ theory in (1 + 1)
dimensions, they were correct, but missed the entire quantum soliton
phenomenon, which is the only physically interesting feature of that
model.

A statement by Yang accurately describes physicists' historical use of
mathematics.

\begin{quote}
``\dots physics is not mathematics, just as mathematics is not
physics.  Somehow nature chooses only a subset of the very beautiful
and complex and intricate mathematics that mathematicians develop, and
that precise subset is what the theoretical physicist is trying to
look for.''
\end{quote}
\vspace{-3ex}\rightline{(Yang)}\vspace{2ex}\noindent
This conservative view of mathematics differs from
Dirac's radical advice, cited earlier, that physicists should

\begin{quote}
``\dots try to interpret \dots mathematical features in terms of
physical entities.''
\end{quote}
\vspace{-3ex}\rightline{(Dirac)}\vspace{2ex}\noindent

However, today faced with the absence of new experimental data about
fundamental phenomena, particle physics theory, as realized in the
string program, is driven by mathematics in the manner advocated by
Dirac.  This was not the way things worked in the past, not even for
Dirac: when first confronting his negative energy solutions, he
identified them with the proton --- the only then-known positively
charged particle --- as a physicist he did not at first trust his
mathematics enough to postulate the existence of the positron!

I am immensely curious about the ultimate fate of the new physics,
built entirely on mathematics, indeed on new mathematics that it helps
to create.  I trust that in the next quarter century the Centre de
Rech\-erches Math\'{e}matiques will play a role in settling this
question.
%}

\end{document}